\documentclass[aps,reprint,pre,10pt,superscriptaddress,nobibnotes,showpacs]{revtex4-1}
\usepackage{amsfonts}
\usepackage{amsmath}
\usepackage{amssymb}
\usepackage{graphicx}%
\usepackage{color}
\begin{document}
\title{Electromagnetic surface modes in a magnetized quantum electron-hole plasma}
\author{A. P. Misra}
\email{apmisra@visva-bharati.ac.in}
\altaffiliation{On leave from Department of Mathematics, Visva-Bharati University, Santiniketan-731 235, India.}
\affiliation{Department of Physics, Ume{\aa } University, SE-901 87 Ume{\aa}, Sweden}


\begin{abstract}
The propagation of surface  electromagnetic waves along a uniform magnetic field is studied in a quantum electron-hole semiconductor plasma. A new forward propagating mode, not reported before, is found by the effect of quantum tunneling, which otherwise does not exist. In the classical limit ($\hbar\rightarrow 0$) one of the low-frequency modes is found similar to an experimentally observed one in $n$-type InSb at room temperature. The surface modes are shown to be significantly modified in the case of high-conductivity semiconductor plasmas where electrons and holes may be degenerate.  The effects of the external magnetic field and the quantum tunneling on the surface wave modes are discussed. 
\end{abstract}
\date{13 April 2011}
\maketitle
The propagation of surface electromagnetic waves along an external magnetic field in conducting solids has been a topic of important research over the last forty years. In fact, there was excellent agreement between the theoretical predictions based on simple models and the experimental observations. In 1972, Baibakov and Datsko \cite{e-h-experiment} had experimentally observed a new low-frequency surface wave mode along a constant magnetic field in  electron-hole (e-h) plasmas  in n-InSb samples at room temperature. This new surface mode was, however, theoretically explained by Flahive and Quinn \cite{e-h-PRL} and later by Uberoi and Rao \cite{e-h-SSC} with the predictions of additional surface modes. In contrast to a single-component plasma, two additional characteristic frequencies, namely  the plasma  and the cyclotron frequencies of holes exist in  e-h plasmas. Thus, new types of  excitations may result into the propagation of surface waves with new interesting behaviors. 

On the other hand, even though the particle number density in semiconductors is lower than that in metals, the high-degree of miniaturization of  today's electronic components opens up the possibility  that the thermal de Broglie wavelength of charge particles may be comparable to or even larger than the spatial variation of the doping profiles. Thus, one could expect the typical quantum mechanical effects, such as tunneling to play important roles in electronic devices to be constructed in near future. A recent review of quantum collective phenomena and typical quantum effects on wave-particle and wave-wave interactions can be found in the literature \cite{review-quantum-plasmas}. Furthermore, in the recent years there has been a considerable interest in the investigation of various surface wave modes in classical (see, e.g. Refs.\cite{surface-semiconductor-classical,surface-classical-PRE,surface-classical-apm,surface-classical-periodic,surface-classical-dusty,surface-bounded,surface-metallic-screen,surface-nonlinear}) as well as in quantum plasmas (see, e.g. Refs.  \cite{quantum-surface-plasma-layer,quantum-surface-plasma-half-space,quantum-surface-e-p,quantum-surface-spin,quantum-surface-Langmuir}).  However, most of these studies until now have been restricted to single-component magnetized or unmagnetized quantum plasmas. 

In this paper, we investigate the propagation of electromagnetic surface waves at the e-h plasma-vacuum interface parallel to an applied magnetic field. We consider the quantum tunneling effect to be associated with the Bohm potential which provides a dispersion due to particle's wave-like nature \cite{review-quantum-plasmas}. In addition to a surface mode, which in the nonretarded limit $(k\gg\omega)$ is given by $\omega\approx\left(1+k/\sqrt{1+k^2+1/m\delta}\right)^{-1/2}$ (in nondimensional form), where $k$ $(\omega)$ is the wave number (frequency), $m$ is the electron to hole mass ratio and $\delta$ is the ratio of the hole to electron number densities, we find other  surface modes  with frequencies below the hole-cyclotron frequency as well as between hole and electron-cyclotron frequencies. Furthermore, a new quantum surface mode is found to exist as a forward propagating wave by the quantum tunneling effect.  In the classical limit, $\hbar\rightarrow 0$, one of the low-frequency modes is found to have similar property with the experimentally observed wave in Ref. \cite{e-h-experiment}. 

We consider a  Cartesian geometry where the plane $x=0$ separates the  half-space $x>0$ filled by a quantum plasma consisting of electrons and holes (to be denoted respectively by $\alpha=e$ and $h$) and vacuum  $(x<0)$. We also assume that the electron and hole densities are, in general, not equal \cite{e-h-unequal}. In a uniform  magnetic field $\mathbf{B}_{0}=B_{0}\hat{z}$, the dynamics of electrons and holes are  governed by \cite{review-quantum-plasmas}
\begin{eqnarray}
& & \partial_t n_{\alpha}+\nabla\cdot\mathbf{v}_{\alpha}=0,\label{1}\\
& & m\frac{\partial\mathbf{v}_{e}}{\partial t}=-\left(\mathbf{E+}\omega_{ch}\mathbf{v}_{e}\times\hat{z}\right)-\kappa\mathbf{\nabla}n_{e}+\frac{H^{2}}{4}\nabla\nabla^{2}n_{e},\label{2}\\
& & \frac{\partial\mathbf{v}_{h}}{\partial t}=\mathbf{E}+\omega_{ch}\mathbf{v}_{h}\times\hat{z}-\sigma\kappa\mathbf{\nabla}n_{h}+\frac{mH^{2}}{4}\nabla\nabla^{2}n_{h},\label{3}
\end{eqnarray}
whereas the electromagnetic wave fields are described by the following Maxwell-Poisson  equations. %
\begin{eqnarray}
& & \nabla\times\mathbf{E}=-\partial_t\mathbf{B},\label{4}\\
& & \nabla\times\mathbf{B}=\mathbf{v}_{h}-{\mathbf{v}_{e}}/{\delta}+\partial_t\mathbf{E},\label{5}\\
& & \nabla\cdot\mathbf{B}=0,\label{6}\\
& & \nabla\cdot\mathbf{E}=n_{h}-{n_{e}}/\delta, \label{7}
\end{eqnarray}
where the number density $n_{\alpha}$ and the velocity $\mathbf{v}_{\alpha}$ for $\alpha$-species particle are   normalized respectively by the equilibrium value $n_{\alpha0}$  and  $c_{s}$. Here $c_{s}\ $may be
defined as either $c_{s}=\sqrt{k_{B}T_{e}/m_{h}}$ for moderate densities
(using, e.g., an isothermal equation of state) or $c_{s}=\sqrt{2k_{B}%
T_{Fe}/m_{h}}$ for relatively dense medium (where electrons and holes are degenerate and  Fermi-Dirac pressure
law pertaining to a three-dimensional zero-temperature Fermi gas is applicable) in which
$T_{F\alpha}$ ($T_{\alpha}$) is the Fermi (thermodynamical) temperature of electrons
or holes with $k_{B}$ denoting the Boltzmann constant. The electric and the magnetic fields $\mathbf{E}$ and  $\mathbf{B}$ are respectively 
 normalized  by $m_{h}c_{s}\omega_{ph}/e$ and $m_{h}c_{s}\omega_{ph}/\sqrt{\epsilon_0\mu_0}e$.
Moreover, the space and the time variables are normalized by $c_{s}%
/\omega_{ph}$ and $\omega_{ph}^{-1}$ respectively. Also, $\omega_{p\alpha
}=\sqrt{n_{\alpha0}e^{2}/\varepsilon_{0}m_{\alpha}}$ is the plasma frequency,
$\omega_{c\alpha}\equiv eB_{0}/m_{\alpha}\omega_{ph}$ is the normalized
cyclotron frequency, $m=m_{e}/m_{h}$ is the electron to hole mass ratio$\ $and
$\delta=n_{h0}/n_{e0}.$ Furthermore, the quantum coupling parameter $H$ appearing in Eqs. (\ref{2}) and (\ref{3}) may be
defined as $H=\sqrt{\delta}\hbar\omega_{pe}/k_{B}T_{e}$ for classical thermal
spread or $H=\sqrt{\delta}\hbar\omega_{pe}/2k_{B}T_{Fe}$ for   degenerate dense plasmas.
In the latter case, which may be relevant for high-conductivity semiconductor plasmas, the temperature ratio will be related to the density ratio
according to $\sigma\equiv T_{Fh}/T_{Fe}=\delta^{2/3}/m,$ where $k_{B}%
T_{F\alpha}=\hbar^{2}(3\pi^{2})^{2/3}n_{\alpha}^{2/3}/2m_{\alpha}$ and
$\hbar\equiv h/2\pi$ is the reduced Plank's constant. Thus, $\kappa=1$
corresponds to classical thermodynamical temperature or moderate density where $\sigma=T_h/T_e$ and $\kappa=1/3$ that for
the Fermi pressure or relatively the dense medium in which electrons and holes are degenerate.

In what follows, we will find solutions that represent the surface waves propagating along the interface $x=0$. To this end, we assume that the electromagnetic fields and the perturbed densities associated with the surface wave with  the wave number $k$ and the  frequency $\omega$ $(<\omega_{ph})$ vary as $\Psi(x,y,t)=$ $\Psi(x)\exp(iky-i\omega t)$. Thus,  we obtain from Eqs. (\ref{1})-(\ref{3}) and (\ref{7})
the wave equations for the  density perturbations as
\begin{equation}
{\partial_x^{2}n_{\alpha}}-\gamma_{\alpha}^{2}n_{\alpha
}=0,\label{8}%
\end{equation}
where
\begin{equation}
\gamma_{\alpha}=\left[\frac{\left(  k^{2}\beta_{2}+\nu_{e}+1/\delta\right)
\left(  k^{2}\beta_{1}+\nu_{h}+1\right)  -1/\delta}{3k^{2}\beta_{1}\beta
_{2}+\left(  \nu_{e}+\mu_{\alpha}\right)  \beta_{1}+\left(  \nu_{h}+1\right)
\beta_{2}}\right]^{1/2}\label{9}%
\end{equation}
with $\beta_{1}=\sigma\kappa+mH^{2}k^{2}/4$,  $\beta_{2}=\kappa+H^{2}k^{2}/4$, $\nu_{e}=m(\omega_{ce}^{2}-\omega^{2})$, $\nu_{h}=\omega_{ch}^{2}-\omega^{2}$,   $\mu_{e}=1/\delta$ and  $\mu_{h}=\delta$.  In obtaining  Eq. (\ref{8}), the very slow nonlocal variations are neglected, i.e.,
$\partial^{6}/\partial x^{6}$, $k^{-2}(\partial^{4}/\partial x^{4})\ll\partial^{2}/\partial x^{2}\ll k^{2}$. In this approximation \cite{quantum-surface-plasma-half-space,quantum-surface-e-p}, the existence of some particular modes, e.g., degenerate or singular waves \cite{singular-wave}, not of current interest, is disregarded.  Next, the  equation for the magnetic field  is
\begin{equation}
{\partial_x^{2}\mathbf{B}}-\alpha_{p}^{2}\mathbf{B}=0,\label{10}%
\end{equation}
where $\alpha_{p}=\left(1+k^{2}-\omega^{2}+1/\delta m\right)^{1/2}$.  The solutions
of Eqs. (\ref{8}) and (\ref{10}) are then given by
\begin{eqnarray}
& & n_{\alpha}=A_{\alpha}\exp(-\gamma_{\alpha}x),\hskip 10pt x>0, \label{11}\\
& & \mathbf{B}=\mathbf{F}_{v}\exp(\alpha_{v}x),\hskip 10pt x<0, \label{12}\\
& & \mathbf{B}=\mathbf{F}_{p}\exp(-\alpha_{p}x), \hskip 10pt x>0, \label{13}
\end{eqnarray}
where  $\alpha_{v}=\sqrt{k^{2}-\omega^{2}}$ is the decay variable of the wave into
vacuum and $A_{\alpha}$, $\mathbf{F}_{v,p}$ are arbitrary constants. Using  the Maxwell equations (\ref{4})-(\ref{6}),  the solutions for the electric field can be obtained as
\begin{align}
\mathbf{E}=&\mathbf{R}_{v}\exp(\alpha_{v}x),\hskip 10pt x<0, \label{14}\\
\mathbf{E}=&\mathbf{R}_{p}e^{-\gamma x}\nonumber\\
& +\sum\limits_{\alpha=e,h}\zeta_{\alpha}A_{\alpha}\Lambda_{\alpha}(-\gamma_{\alpha}\hat{x}+ik\hat{y})e^{-\gamma_{\alpha}x},\hskip 5pt x>0, \label{15}%
\end{align}
where $\mathbf{R}_{v,p}$ are arbitrary constants with $\zeta_{e,h}=\mp1$ and
\begin{align}
\gamma=&\left[k^{2}-\omega^{2}-\frac{\omega^{2}}{\omega_{ch}^{2}-\omega^{2}}-\frac{\omega^{2}}{m\delta(\omega_{ce}^{2}-\omega^{2})}\right]^{1/2}, \label{16}\\
\Lambda_{e}=&\left[  1+\frac{m\omega^{2}}{\omega_{ch}^{2}-\omega^{2}}\left(\kappa+\frac{H^{2}}4\left(\gamma_{e}^{2}-k^{2}\right)\right)\right]\nonumber\\
&\times \left[{1}/{\delta(\gamma_{e}^{2}-\gamma^{2})}\right], \label{17}\\
\Lambda_{h}=&\left[1+\frac{\omega^{2}}{\omega_{ch}^{2}-\omega^{2}}\left(\kappa{\sigma}-\frac{mH^{2}}4\left(\gamma_{h}^{2}-k^{2}\right)\right)\right]\nonumber\\
&\times\left[{1}/{\left(\gamma_{h}^{2}-\gamma^{2}\right)}\right]. \label{18} 
\end{align} 
\begin{figure}[btp]
\includegraphics[width=3in,height=2.1in,trim=0.0in 0in 0in 0in]{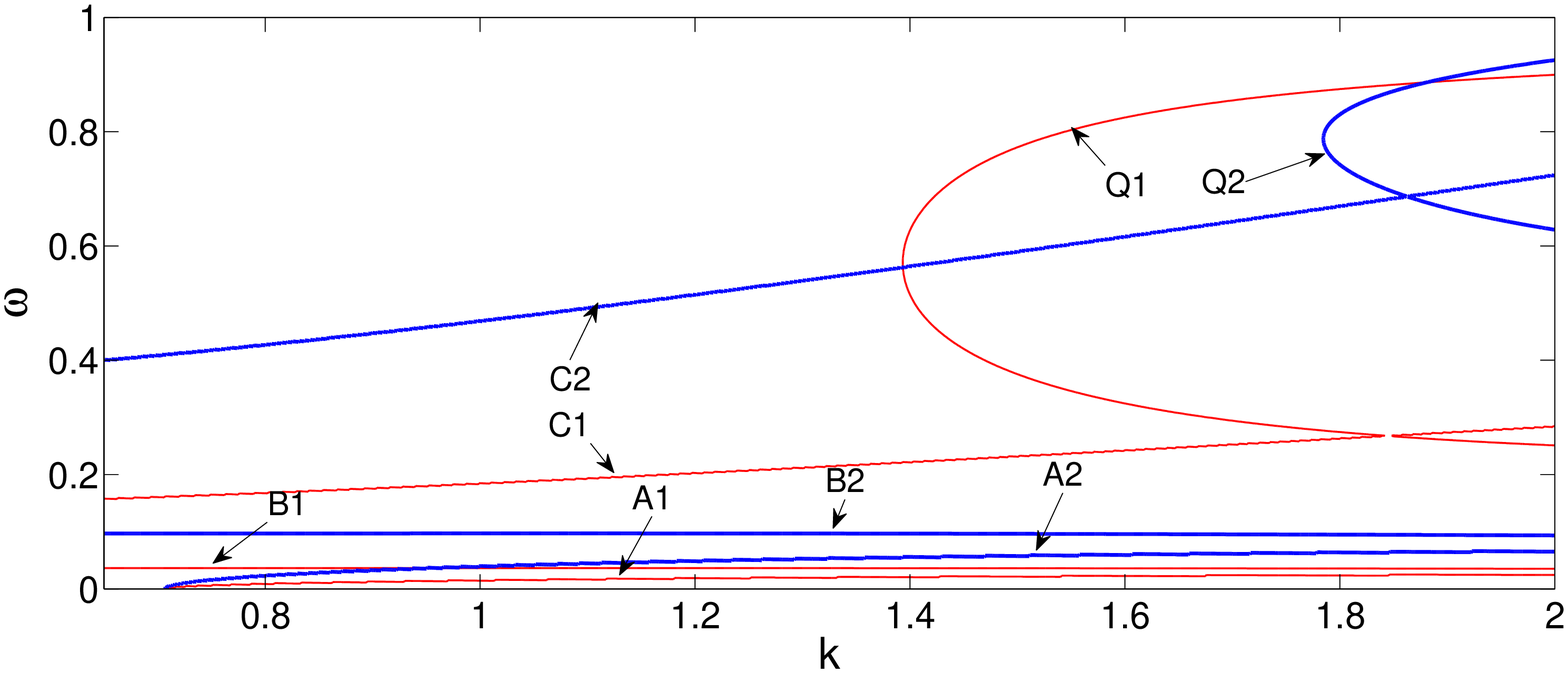} \caption{ (Color online) The dispersion relation (\ref{19}) in non-dimensional form is contour plotted to show different surface wave modes in the $k-\omega$ plane (The case of non-degenerate plasmas). The modes labeled $A$'s, $B$'s, $C$'s are for $H=0$ and those with labels  $Q$'s are for nonzero $H$. The labels [$A1$, $B1$, $C1$, $Q1$; the thin (red) lines] and [$A2$, $B2$, $C2$, $Q2$; the thick (blue) lines] indicate the surface modes corresponding to $B_0=0.6$ T and $B_0=1.6$ T respectively.  Other parameters are $m_e=0.01m_0$, $m_h=0.4m_0$, $T_e=T_h\approx300$ K, $n_{e0}=1.35\times10^{22}$ m$^{-3}$  and $n_{h0}=10^{22}$ m$^{-3}$.}
\end{figure}
\begin{figure}[btp]
\includegraphics[width=3in,height=2.1in,trim=0.0in 0in 0in 0in]{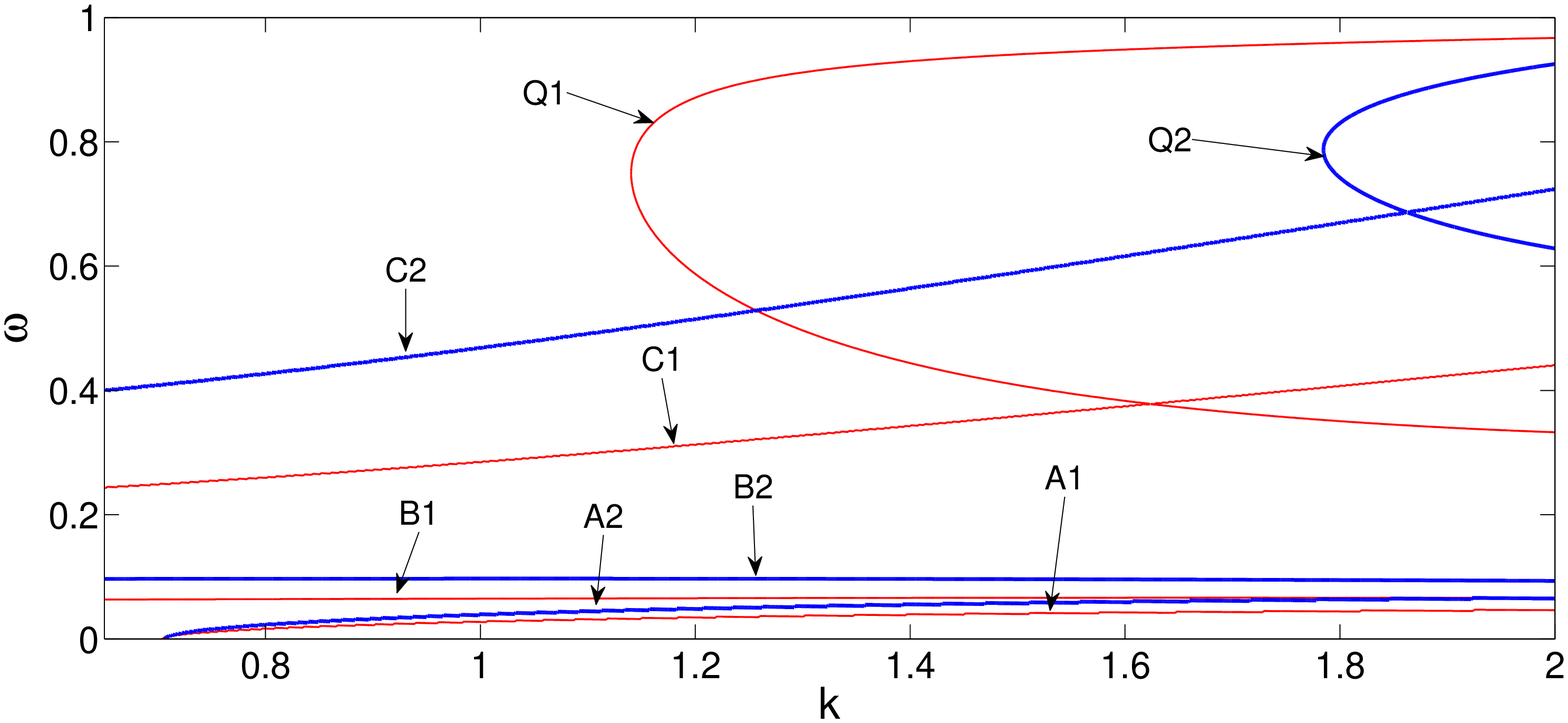} \caption{ (Color online) The same as in Fig. 1, but for the effect of density variation. The modes labeled $A$'s, $B$'s, $C$'s are for $H=0$ and those with labels  $Q$'s are for nonzero $H$. The labels [$A1$, $B1$, $C1$, $Q1$; the thin (red) lines] and [$A2$, $B2$, $C2$, $Q2$; the thick (blue) lines] indicate the surface modes corresponding to the densities $n_{e0}=4\times10^{22}$ m$^{-3}$, $n_{h0}=2\times10^{22}$ m$^{-3}$  and $n_{e0}=1.35\times10^{22}$ m$^{-3}$, $n_{h0}=10^{22}$ m$^{-3}$ respectively.  Other parameters are the same as for the thick (blue) lines in Fig. 1.}
\end{figure}
Surface waves are those solutions for which the wave number normal to the surface has negative imaginary part, leading to an exponential decay away from the surface. Thus, in the above solutions (\ref{11})-(\ref{15}) we have retained  only those parts  in both the regions which exponentially decay away from the interface.   Next, we use the  boundary conditions, namely (i) the tangential component of $\mathbf{E}$ and $\mathbf{B}$ are continuous at $x=0$, (ii) the normal component of the displacement vector is continuous at $x=0$,  (iii)  velocity components (along $x$ axis) vanish for both electrons and holes (hot species), i.e. $\mathbf{v}_{ex}=\mathbf{v}_{hx}=0$ at $x=0$. Thus, we obtain a system of linear homogeneous equations which has nontrivial solutions only if the determinant of the resulting system vanishes. This leads to the following dispersion relation.
\begin{align}
 \left[\omega^{2}\alpha_{v}+\left(\omega^{2}-1\right)\alpha_{p}\right]\left[\frac{\left[  \omega^{2}\gamma_{e}-\left(\omega^{2}-1\right) \alpha_{p}\right]  \Delta_{e}}{\delta(\gamma_{e}^{2}-\gamma^{2})(\beta_{2}-H^{2}\gamma_{e}^{2}/4)}\right.  \nonumber\\
\left. +(2\omega^{2}-1)\alpha_{v} +\frac{\left[  \omega^{2}\gamma_{h}-\left(  \omega^{2}-1\right)  \alpha_{p}\right]  \Delta_{h}}{\left(\gamma_{h}^{2}-\gamma^{2}\right)(\beta_{1}-mH^{2}\gamma_{h}^{2}/4)}\right]=0,\label{18}%
\end{align}
where
\begin{align}
&\Delta_{e}=1+\frac{\omega^{2}}{m\left(  \omega_{ce}^{2}-\omega^{2}\right)}\left(  \kappa-\frac{H^{2}k^{2}}{4}\right)  +\frac{H^{2}\gamma_{e}^{2}}{4},\label{19}\\
&\Delta_{h}=1+\frac{\omega^{2}}{\omega_{ch}^{2}-\omega^{2}}\left(\sigma\kappa+\frac{mH^{2}k^{2}}{4}\right)-\frac{mH^{2}\gamma_{h}^{2}}4.\label{20}%
\end{align}
\begin{figure}[btp]
\includegraphics[width=3in,height=2.1in,trim=0.0in 0in 0in 0in]{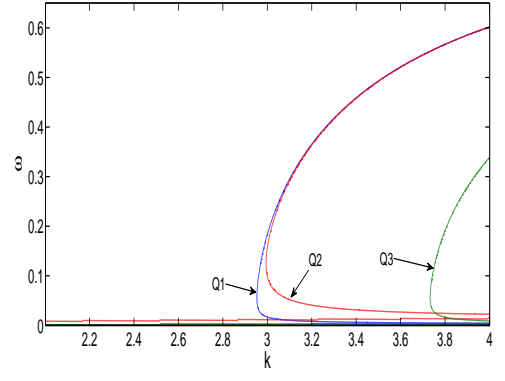} \caption{ (Color online) The dispersion relation (\ref{19}) in non-dimensional form is contour plotted to show different surface wave modes in the $k-\omega$ plane (The case of degenerate dense plasmas). The effects of the variations of both the magnetic field and the particle number density are shown. One low-frequency $(<\omega_{ch})$ mode appears (not labeled) whose slope increases with the magnetic field and decreases with the density enhancement. The lines labeled $Q1$ and $Q2$  are the quantum surface  modes corresponding to $B_0=0.6$ T and $B_0=3$ T respectively with fixed  $m_e=0.02m_0$, $m_h=0.4m_0$,  $n_{e0}=2\times10^{26}$ m$^{-3}$  and $n_{h0}=10^{26}$ m$^{-3}$. The quantum mode (labeled $Q3$) is for $B_0=0.6$ T,  $n_{e0}=3\times10^{26}$ m$^{-3}$, $m_e=0.02m_0$ and $m_h=0.4m_0$.}
\end{figure}
We note that the surface waves occur only if $k>\omega$. The first factor of the dispersion equation (\ref{18}) gives ordinary surface mode (as mentioned in the introduction) independent of the magnetic field and the quantum correction term, which decays with the wave number $k$. However, it approaches a constant value $\omega\approx0.7$ in the limit $k\gg\omega$. In order to analyze numerically the dispersion relation (\ref{19}), we consider two different density regimes relevant for non-degenerate and degenerate  plasmas. We consider the fixed parameters $m_e=0.01m_0$, $m_h=0.4m_0$, $T_e=T_h\approx300$ K, where $m_0$ is the free electron mass. For non-degenerate particles (Figs. 1 and 2),  we use the densities $n_{\alpha0}\sim10^{22}$ m$^{-3}$ as in Refs. \cite{e-h-experiment,e-h-PRL} and for degenerate electrons and holes we consider $n_{\alpha0}\sim10^{26}$ m$^{-3}$ (Fig. 3). In the latter,  the thermal de Broglie wavelength $\lambda_B$ is greater than the average inter-particle distance, i.e. $n_{\alpha0}\lambda^3_B\sim17>1$, and so quantum effect is no longer negligible. Thus, the typical quantum mechanical effects, e.g. tunneling will certainly play an important role in the modification and/ or generation of a new dispersive surface mode. Notice that the quantum modified modes may not appear in the present case as we have disregarded the very slow nonlocal variations \cite{quantum-surface-plasma-half-space,quantum-surface-e-p}, rather appears a new mode by the quantum force, which otherwise (i.e. in the limit $\hbar\rightarrow0$) does not exist.

In  Figs. 1-3, the lines with labels $A$'s, $B$'s and $C$'s correspond to the classical ($H=0$) surface modes whereas those with labels $Q$'s are due to quantum tunneling effects.  We find that there appear three different modes for $H=0$ whose frequencies lie in the regimes: $\omega_{ch}<\omega<\omega_{ce}$, $\omega_{ch}<\omega\approx\omega_1<\omega_{ce}$ and $0<\omega<\omega_{ch}$, where $\omega_1$ depends on the parameters to be chosen. In the latter, the low-frequency mode labeled $A1$ or $A2$  has the similar behaviors as experimentally observed in Ref. \cite{e-h-experiment}.  Its slope increases with increasing the strength of the magnetic field and the wave phase speed tends to zero at a nonzero $k=k_1$, i.e. the existence of surface wave at $k<k_1$ is not possible. This particular mode was theoretically explained by Flahive et al \cite{e-h-PRL} and later by Uberoi et al \cite{e-h-SSC}, however, there was no quantitative verification due to lack of experimental data in Ref. \cite{e-h-experiment}. In Fig. 1, the surface modes corresponding to the magnetic field $B_0=1.6$ T appear as thick solid lines, and the thin lines are due to $B_0=0.6$ T. From Fig. 1, we also find that as the magnetic field strength increases, the forward propagating quantum surface mode labeled $Q1$ or $Q2$ appears at higher values of $k$ and the frequency domain in which it appears reduces.  In contrast to Fig. 1, Fig. 2 shows that the slopes of the classical modes decrease with  increasing the electron to hole density ratio or corresponding to a higher density regime ($\lesssim10^{24}$ m$^{-3}$). Also, the quantum mode appears in a larger frequency domain with smaller wave numbers.   

Next, we consider the case in which electrons and holes are degenerate. This may be relevant for high-conductivity semiconductors where the particle number density may exceed, e.g. $10^{24}$ m$^{-3}$.   In this case, the pressures of degenerate electrons and holes can be described by the Fermi-Dirac pressure law. Thus, we consider the parameters $m_e=0.02m_0$, $m_h=0.4m_0$, the densities $n_{\alpha0}\sim10^{26}$ m$^{-3}$ and the magnetic field $B_0\sim1$ T. We see from Fig. 3 that one low-frequency mode appears which approaches a constant value $(<\omega_{ch})$ at higher values of $k$. Note that the classical mode can not be recovered in this case for $H=0$ as the denominator of $H$ will now be the Fermi energy. Furthermore, $H\rightarrow0$ corresponds to extremely dense regimes which might not be relevant for semiconductor plasmas.  For the quantum modes (labeled $Q1$, $Q2$ and $Q3$), there exists a critical value of the wave frequency below which the phase speed increases with increasing the magnetic field and above which it remains 
unchanged (lines labeled $Q1$ and $Q2$). In contrast to Fig. 2 (the case of non-degenerate particles), the enhancement of the electron density (keeping other parameters fixed) shows  that the quantum mode appears in a smaller frequency domain at higher values of $K$. 

To summarize, a new quantum surface mode at a plasma-vacuum interface is shown to appear by the inclusion of quantum mechanical (tunneling) effect in a magnetized electron-hole semiconductor plasma. Recent technological progress in the creation of smaller scales of plasma oscillations in electronic devices indicate that the thermal de Broglie wavelength can even be larger than the inter-particle distance of the charge carriers, and so quantum mechanical effects (tunneling) may no longer be neglected.  We note that such  surface mode due to quantum tunneling  appears as a forward propagating wave upto a certain frequency below the hole plasma frequency. Furthermore, the phase speed of these waves strongly depends on the external magnetic field and the electron-hole concentration in the plasma. Apart from that several other low-frequency modes  also appear in the limit $\hbar\rightarrow0$, one of which has similar behaviors with that experimentally observed mode in n-InSb at room temperature \cite{e-h-experiment}.  On the other hand,  when quantum statistical effects are taken into consideration along with the quantum tunneling, specifically for high-conductivity semiconductors where electrons and holes are rather dense and may be degenerate, the quantum surface wave propagates in a different way in contrast to non-degenerate plasmas. The results may be useful for understanding the dispersion properties of new quantum surface waves in semiconductor plasmas, which can be   observed experimentally in near future. 
 
{This work was supported by the Kempe Foundations, Sweden through Grant No. SMK-2647. Author wishes to thank S. Samanta of Department of Basic Science and Humanities, College of Engineering and Management, India for his  help. Thanks are due to Gert Brodin and Mattias Marklund of Department of Physics, Ume\aa\ University, SE-901 87 Ume\aa, Sweden for  support.}


\begin{thebibliography}{50}
\bibitem{e-h-experiment} V. I. Baibakov and N. Datskov, Sov. Phys. JETP Lett. \textbf{15}, 135 (1972).
\bibitem{e-h-PRL} P. G. Flahive and J. J. Quinn, Phys. Rev. Lett. \textbf{31}, 586 (1973).
\bibitem{e-h-SSC} C. Uberoi and U. J. Rao, Solid State Commu. \textbf{21}, 579 (1977).
\bibitem{review-quantum-plasmas} P. K. Shukla and B. Eliasson, Phys. Usp. \textbf{53}, 51 (2010).
\bibitem{surface-semiconductor-classical} C. R. Legendy, Phys. Rev. A \textbf{135}, 1713 (1964).
\bibitem{surface-classical-PRE} A. P. Misra and A. R. Chowdhury, Phys. Rev. E \textbf{70}, 058401 (2004).
\bibitem{surface-classical-apm} A. P. Misra and A. R. Chowdhury, Phys. Scr. \textbf{69}, 44 (2004).
\bibitem{surface-classical-periodic} P. K. Shukla and L. Stenflo, Phys. Plasmas \textbf{12}, 044503 (2005).
\bibitem{surface-classical-dusty} R. Bharuthram and P. K. Shukla, Planet. Space Sci.  \textbf{41}, 17 (1993).
\bibitem{surface-bounded} O. M. Gradov and L. Stenflo, Phys. Rep. \textbf{94}, 111 (1983).
\bibitem{surface-metallic-screen} K. S. Yi \textit{et al}, Phys. Rev. B \textbf{22}, 6247 (1980).
\bibitem{surface-nonlinear} O. M. Gradov and L. Stenflo, J. Plasma Phys.  \textbf{65}, 73 (2001).
\bibitem{quantum-surface-plasma-layer} B. Shokri and A. A. Rukhadze, Phys. Plasmas \textbf{6}, 3450 (1999).
\bibitem{quantum-surface-plasma-half-space} M. Lazar, P. K. Shukla and A. Smolyakov, Phys. Plasmas \textbf{14}, 124501 (2007).
\bibitem{quantum-surface-e-p} A. P. Misra, N. K. Ghosh and P. K. Shukla, J. Plasma Phys. \textbf{76}, 87 (2010).
\bibitem{quantum-surface-spin} A. P. Misra, Phys. Plasmas \textbf{14}, 064501 (2007).
\bibitem{quantum-surface-Langmuir} I. S. Chang and D. Y. Jung, Phys. Lett. A \textbf{9}, 372 (2008).
\bibitem{e-h-unequal} At room temperatures the conductivity of intrinsic semiconductors in which electron and hole densities are equal may be relatively  low. However, this can be enhanced by adding some  impurities (doping) in the background plasma. Furthermore, in the case of $n$-type (extrinsic) semiconductors which were also used in the experiment of Ref. \cite{e-h-experiment} as a sample, the electrons are considered to be major carriers.   
\bibitem{singular-wave} A. P. Misra and S. Samanta, Phys. Plasmas \textbf{16}, 074505 (2009).
\end{thebibliography}
\end{document}